\documentstyle[pra,aps]{revtex} 

\tighten

\newcommand{\bra}[1]{\langle #1|}
\newcommand{\ket}[1]{|#1\rangle}
\newcommand{\braket}[2]{\langle #1|#2\rangle}
\newcommand{\ketbra}[2]{|#1\rangle\langle #2|}

\newcommand{\Exp}[1]{\mathrm{e}^{\mbox{\footnotesize$#1$}}}
\newcommand{\pow}[1]{^{\mbox{\footnotesize$#1$}}}
\newcommand{\visib}{{{\mathcal{V}}}}
\newcommand{\trenv}[1]{{\rm tr}_{\rm env}\left\{#1\right\}}

\begin{document}

\draft
\title{On mechanisms that enforce complementarity}
\author{Berthold-Georg Englert,$^{\ast\dagger}$
Marlan O. Scully,$^{\ast\ddagger}$  
and Herbert Walther$^{\ast\S}$}
\address{%
$^\ast$Max-Planck-Institut f\"ur Quantenoptik, 
Hans-Kopfermann-Strasse 1, 85748 Garching, Germany\protect\\
$^\dagger$Abteilung f\"ur Quantenphysik, Universit\"at Ulm,
Albert-Einstein-Allee~1, 89081 Ulm, Germany\protect\\
$^\ddagger$Department of Physics, Texas A\&M University, College Station,
TX 77843-4242, USA\protect\\
$^\S$Sektion Physik, Universit\"at M\"unchen, 
Am Coulombwall 1, 85748 Garching, Germany}
\date{8 October 1999}

\ifpreprintsty\relax\else\wideabs{\fi%
\maketitle
\begin{abstract}\ifpreprintsty\relax\else\hspace*{1em}\fi%
In a recent publication Luis and S\'anchez-Soto arrive at the conclusion that
complementarity is universally enforced by random classical phase kicks. 
We disagree. One could just as well argue that quantum entanglement is the
universal mechanism. Both claims of universality are unjustified, however. 
\end{abstract}
\ifpreprintsty\relax\else}\fi

\narrowtext

\section{Introduction and summary}
\begin{mathletters}\label{eq:0}
In an interesting recent paper \cite{LSS}, Luis and S\'anchez-Soto
reconsidered the loss of fringe visibility in an interferometer with which-way
detection devices, and they conclude that this loss occurs because
\begin{equation}\label{eq:0a}
\parbox{0.7\columnwidth}{\raggedright%
random classical phase kicks enforce complementarity \emph{always}.}
\end{equation}
We disagree.
In particular, the classical-random-phase picture of Ref.\ \cite{LSS} does not 
apply to the thought experiment of a double-slit interferometer for atoms with
quantum-optical which-way detection that we introduced in Ref.~\cite{SEW}
and treated anew in Ref.~\cite{ESWajp}.
Under the idealized circumstances of a thought experiment,
the dynamics is absolutely free of noise and therefore perfectly unitary,
and there is no room for any element of \emph{classical} randomness.

As we shall show here, one could just as well claim that an
opposite of (\ref{eq:0a}) is true, namely that
\begin{equation}\label{eq:0b}
\parbox{0.7\columnwidth}{\raggedright%
quantum entanglement enforces complementarity \emph{invariably}.}
\end{equation}
In this generality, statements (\ref{eq:0a}) and (\ref{eq:0b}) are both
excessive, although one could argue,
as we show below, that they are mathematically correct in the sense that
there are formal procedures that make matters appear either way.
There are situations in which (\ref{eq:0a}) or (\ref{eq:0b}), or a
combination of both, gives an accurate account of the actual physical
mechanism that reduces the fringe visibility, but there is no justification
for the universality associated with ``always'' or ``invariably.'' 
\end{mathletters}

Rather than ``enforce(s) complementarity'' one should say more precisely
``lead(s) to a loss of fringe visibility'' in (\ref{eq:0a}) and
(\ref{eq:0b}).
Complementarity becomes an issue only when which-way information is made
available and wave-particle duality requires a corresponding reduction of the
fringe visibility.
If truly classical random phase kicks are at work, fringes get lost but
nothing can be learned in return about the way through the interferometer,
and so the question of how complementarity is enforced doesn't come up.
Establishing quantum entanglement is the first step in acquiring which-way
knowledge, and then the fringe visibility is accordingly reduced because the
``quantum aspects of the which-way detection enforce duality and thus make
sure that the principle of complementarity is not circumvented''
\cite{BGEprl96}.

\section{Concerning the thought experiment of Ref.\ \protect\cite{SEW}}
Let us say a few words about the thought experiment of Ref.\ \cite{SEW} before
turning to the general, and more abstract, discussion. 
The crucial quantity is the inner product of $\ket{A_1}$ and $\ket{A_2}$,
specified in Eq.\ (17) of Ref.~\cite{LSS}, viz.\
\begin{equation}\label{eq:x1}
\braket{A_1}{A_2}=\cos^2(\lambda t)\braket{0_10_2}{0_10_2}
+\sin^2(\lambda t)\braket{1_10_2}{0_11_2}\,,
\end{equation}
where $\lambda t$ characterizes the effectiveness of the atom-photon
interaction and $\ket{0_11_2}$, for example, is the state with no photons in
cavity $C_1$ of Fig.~2 in Ref.~\cite{LSS} and one photon in cavity $C_2$.
Since $\ket{1_10_2}$ and $\ket{0_11_2}$ are orthogonal photon states, we have
$\braket{A_1}{A_2}=\cos^2(\lambda t)$, of course, and this can be
written in the form 
\begin{equation}\label{eq:x2}
\braket{A_1}{A_2}=\int\!d\phi\,\Exp{i\phi}\Omega(\phi)\,,
\end{equation}
which seems to suggest an interpretation as a ``phase randomization'' in the
spirit of Ref.~\cite{LSS}. 

It is true that one can always find a positive weight $\Omega(\phi)$
with all the right properties 
--- the simple choice
\begin{equation}\label{eq:x3}
\Omega(\phi)=\frac{1}{2\pi}+\frac{1}{\pi}\cos\phi\,\cos^2(\lambda t)
\end{equation}
serves the purpose if $\cos^2(\lambda t)\leq\frac{1}{2}$ --- 
but the arbitrarily introduced phase variable $\phi$ cannot have any
physical significance because, as we said earlier, the idealized thought
experiment is free of noise.
In fact, as we'll recall in Sec.\ \ref{sec:QE} below, one can ``undo'' or
reverse the effects of the which-way measurement via quantum erasure, and this
is an important difference between the ``classical random phase kicks'' and
the ``quantum entanglement'' points of view.
If classical phase noise were at work, quantum erasure couldn't be done.

To emphasize the arbitrariness of the weight $\Omega(\phi)$ in (\ref{eq:x2})
let us show how the overcompleteness of the coherent states $\ket{\alpha}$
\cite{SZbook} (and of states derived from them) can be used to write the key
inner-product $\braket{1_10_2}{0_11_2}=0$ of the thought experiment of
\cite{SEW} very easily in the phase-integral form of Ref.\ \cite{LSS}.
We note the completeness relation (one of many)~\cite{FNx}
\begin{equation}\label{eq:x4}
\openone=\int\!\!\frac{d\varphi}{2\pi}\int\!\!dI
\left[\left(a^{\dagger}a\right)!\,I\pow{-a^{\dagger}a}\right]^{1/2}
\ketbra{\alpha}{\alpha}
\left[\left(a^{\dagger}a\right)!\,I\pow{-a^{\dagger}a}\right]^{1/2}
\end{equation}
where $\alpha=\sqrt{I}\exp(i\varphi)$ relates the amplitude $\alpha$ of
the coherent state to the integration variables $I$ and $\varphi$, and $a$,
$a^{\dagger}$ are the ladder operators of the photon mode in question
($a\ket{\alpha}=\ket{\alpha}\alpha$).
Upon using (\ref{eq:x4}) twice and integrating over the intensities $I_1$
and $I_2$ we get
\begin{eqnarray}
\bra{1_10_2}\openone_1\openone_2\ket{0_11_2}&=&
\int\!\!\frac{d\varphi_1}{2\pi}\int\!\!\frac{d\varphi_2}{2\pi}
\Exp{i(\varphi_1-\varphi_2)}\nonumber\\
&=&\int\!\!d\phi\,\Exp{i\phi}\frac{1}{2\pi}\,,
\label{eq:x5}
\end{eqnarray}
where $\phi=\varphi_1-\varphi_2$ in the last step. Without much effort we
have thus arrived at one of the many ways of writing $0$ as a phase integral.
In view of the obvious arbitrariness of the whole procedure, any physical
interpretation of the phase variable $\phi$ is equally arbitrary. 
Therefore, one should not associate ``random classical phase kicks'' with the
$\phi$ integral. 

\section{General discussion}
\subsection{Setting the stage}
We turn to the general discussion now.
As in \cite{LSS}, we proceed from assuming that, when no which-way detection
is attempted, the interferometer prepares the interfering
quantum system (photon, electron, atom, \dots) in the superposition state
$(0<\theta<\pi) $
\begin{equation}\label{eq:1}
\ket{\psi_0}=\cos(\theta/2)\ket{\psi_1}
+\Exp{i\phi}\sin(\theta/2)\ket{\psi_2}\,,
\end{equation}
where $\ket{\psi_1}$ and $\ket{\psi_2}$ symbolize the basic alternatives, 
i.e.\ the \emph{ways}, of the interferometer --- such as ``through this
slit'' and ``through that slit'' in a Young's double-slit experiment.
A measurement of the probability for finding the system in the $\varphi$
dependent state
\begin{equation}\label{eq:2}
  \ket{\psi(\varphi)}=\frac{1}{\sqrt{2}}\left(\ket{\psi_1}
                     +\Exp{i\varphi}\ket{\psi_2}\right)
\end{equation}
produces the interference pattern 
\begin{equation}\label{eq:3}
p_0(\varphi)=\biglb|\braket{\psi(\varphi)}{\psi_0}\bigrb|^2=
\frac{1}{2}\left[1+\sin\theta\,\cos(\varphi-\phi)\right]\,,
\end{equation}
so that $\visib_0=\sin\theta$ is the fringe visibility.

Now, an attempt to gain which-way knowledge is made (or any other
manipulation of the interferometric setup). 
In concordance with Ref.\ \cite{LSS}, we shall take for granted that
the probabilities for finding the basic states $\ket{\psi_1}$ and  
$\ket{\psi_2}$ are not affected. Therefore, the net effect can only be a
change
of the interference pattern from $p_0(\varphi)$ of (\ref{eq:3}) to
\begin{equation}\label{eq:4}
p_\epsilon(\varphi)=
\frac{1}{2}\left[1+|\epsilon|\sin\theta\,
\cos(\varphi-\phi-\delta_\epsilon)\right]\,,  
\end{equation}
where the complex parameter $\epsilon=|\epsilon|\exp(i\delta_\epsilon)$ 
is subject to the restriction 
$|\epsilon|\leq1$. As a consequence, the fringe pattern is shifted
by $\delta_\epsilon$ and its visibility is reduced to 
$\visib=|\epsilon|\visib_0$.

The transition from (\ref{eq:3}) to (\ref{eq:4}) means that the statistical
operator of the quantum system has undergone the transformation
\begin{eqnarray}
\rho_0&=&\ketbra{\psi_0}{\psi_0}\nonumber\\\to\rho_\epsilon&=&
\ket{\psi_1}\cos^2(\theta/2)\bra{\psi_1}
+\ket{\psi_2}\sin^2(\theta/2)\bra{\psi_2}
\nonumber\\&&+
\ket{\psi_1}\cos(\theta/2)\Exp{-i\phi}\epsilon^*\sin(\theta/2)\bra{\psi_2}
\nonumber\\&&+
\ket{\psi_2}\sin(\theta/2)\Exp{i\phi}\epsilon\cos(\theta/2)\bra{\psi_1}\,,  
\label{eq:5}
\end{eqnarray}
which is nonunitary if $|\epsilon|<1$.

\subsection{The ``classical random phase kicks'' argument}
The stage is now set for a  presentation of the ``classical random phase 
kicks'' argument of Ref.\ \cite{LSS}. It has essentially two ingredients. 
{\bf(i)} First,
an actual classical randomization of the phase $\phi$ in (\ref{eq:1}), 
such that additional phase factors $\exp(i\phi')$ with $\phi'$ in the range 
$\phi'\cdots\phi'+d\phi'$ occur with probability $d\phi'\,\Omega(\phi')$,
would cause the transition (\ref{eq:5}) with $\epsilon$  given by \cite{FN1}
\begin{equation}\label{eq:6}
\epsilon=\int\!\!d\phi'\,\Omega(\phi')\Exp{i\phi'}\,,  
\end{equation}
provided that the positive weight $\Omega(\phi')$ is normalized in 
accordance with
\begin{equation}\label{eq:7}
\int\!\!d\phi'\,\Omega(\phi')=1\,.
\end{equation}
{\bf(ii)} Second, any given value of $\epsilon$, in particular the one 
needed 
in (\ref{eq:4}), can be written in the form (\ref{eq:6}) with a suitably
chosen
$\Omega(\phi')$. There is, of course, an abundance of permissible 
$\Omega$-weights, of which
\begin{equation}\label{eq:8}
  \Omega(\phi')=\frac{1+|\epsilon|}{2}\delta(\phi'-\delta_\epsilon)+
\frac{1-|\epsilon|}{2}\delta(\phi'-\delta_\epsilon-\pi)
\end{equation}
and
\begin{equation}\label{eq:8'}
\Omega(\phi')=\left\{
\begin{array}{cl}
(2\chi)^{-1}&\mbox{for $|\phi'-\delta_\epsilon|\leq\chi\leq\pi$}\\&
\mbox{with $|\epsilon|=\chi^{-1}\sin\chi$,}\\[1ex]
0&\mbox{elsewhere,}
\end{array}\right.
\end{equation}
are perhaps the simplest ones; another explicit example is given in Eq.\ (12)
of \cite{LSS}.

Since there is always a $\Omega$-weight with the desired properties, one can
justifiably state that the net effect is always \emph{as if} a classical
randomization of the phase $\phi$ of (\ref{eq:1}) had occurred. 
According to Ref.\ \cite{LSS}, however, this is the \emph{actual} mechanism, 
notwithstanding the said \emph{as-if} character and irrespective of what is
really going on. 

We disagree with this \emph{physical} interpretation of the
\emph{mathematical} fact (ii) --- and thus with the central message of
Ref.~\cite{LSS}.  
The \emph{as-if} aspect is relevant and deserves emphasis.

\subsection{The ``quantum entanglement'' argument}
Before offering further remarks, it is expedient to put the ``quantum 
entanglement'' argument on record, which has also two logical ingredients. 
{\bf(i')} First, an interaction of the system with its environment (part of 
which might be a which-way detector controlled by the experimenter) 
establishes an entangled state \cite{FN2} 
\begin{eqnarray}
&&\Bigl(\ket{\psi_1}\cos^2(\theta/2)\bra{\psi_1}\Bigr)\otimes
U_{\rm env}\rho_{\rm env}U^{\dagger}_{\rm env}
\nonumber\\&+&
\Bigl(\ket{\psi_2}\sin^2(\theta/2)\bra{\psi_2}\Bigr)\otimes\rho_{\rm env}
\nonumber\\&+&\Bigl(
\ket{\psi_1}\cos(\theta/2)\Exp{-i\phi}\sin(\theta/2)\bra{\psi_2}
\Bigr)\otimes \rho_{\rm env}U^{\dagger}_{\rm env}
\nonumber\\&+&\Bigl(
\ket{\psi_2}\sin(\theta/2)\Exp{i\phi}\cos(\theta/2)\bra{\psi_1}  
\Bigr)\otimes U_{\rm env}\rho_{\rm env}
\label{eq:12}  
\end{eqnarray}
and consequently effects a transition of the form (\ref{eq:5}) with
\begin{equation}\label{eq:9}
\epsilon=\trenv{U_{\rm env}\rho_{\rm env}}\,,
\end{equation}
where $\rho_{\rm env}$ is a statistical operator of the environment prior to
the interaction, $U_{\rm env}$ is a unitary operator acting solely on the
environment variables, and $\trenv{\cdots}$ is the injunction to trace over
these variables. 
{\bf(ii')} Second, any given value of $\epsilon$, in particular the one needed
in (\ref{eq:4}), can be written in the form (\ref{eq:9}) with suitably chosen
$\rho_{\rm env}$ and $U_{\rm env}$. 
Here, too, we have an abundance of possible choices. A specific example is a
harmonic oscillator with 
\begin{equation}\label{eq:10}
\rho_{\rm env}=\ketbra{0}{0}
\end{equation}
and $U_{\rm env}$ such that
\begin{eqnarray}
U_{\rm env}\ket{0}&=&\epsilon\ket{0}+\Exp{i\delta_\epsilon}
\sqrt{1-|\epsilon|^2}\,\ket{1}\,,
\nonumber\\
U_{\rm env}\ket{1}&=&\Exp{i\delta_\epsilon}
\sqrt{1-|\epsilon|^2}\,\ket{0}-\epsilon\ket{1}\,,
\label{eq:11}
\end{eqnarray}
where $\ket{0}$ and $\ket{1}$ are the ground and first excited oscillator 
states, respectively.

Since there is always a pair $\rho_{\rm env},U_{\rm env}$ with the desired 
properties, one can justifiably state that it is always \emph{as if} 
a quantum entanglement had caused the transition (\ref{eq:5}). 
But, of course, one cannot disregard its \emph{as-if} character and conclude
that this mechanism  is invariably the \emph{actual} one, irrespective of what
is really going on. 

\subsection{SAI equivalence}
In Ref.~\cite{SAI}, Stern, Aharonov, and Imry (SAI) largely anticipated the
formal aspects of Ref.~\cite{LSS} without, however, arriving at the
questionable statement (\ref{eq:0a}). 
Moreover, SAI have noted that one can turn the
``quantum entanglement'' picture into the ``classical random phase kicks'' 
one by summing over the eigenstates of $U_{\rm env}$ when evaluating
the trace in (\ref{eq:9}). Likewise, for a given $\Omega(\phi')$ one can
always construct a corresponding  $\rho_{\rm env},U_{\rm env}$ pair. The
two \emph{as-if} descriptions are, so to say, ``SAI equivalent.'' In the
example of Eqs.\ (\ref{eq:10}) and (\ref{eq:11}) we have
\begin{equation}\label{eq:13}
U_{\rm env}\ket{\pm}=\pm\Exp{i\delta_\epsilon}\ket{\pm}  
\end{equation}
with
\begin{equation}\label{eq:14}
\ket{\pm}=\sqrt{(1\pm|\epsilon|)/2}\,\ket{0}
\pm\sqrt{(1\mp|\epsilon|)/2}\,\ket{1}\,,
\end{equation}
so that the phases $\delta_\epsilon$ and $\delta_\epsilon+\pi$ have the
respective weights $(1+|\epsilon|)/2$ and $(1-|\epsilon|)/2$, and the
$\Omega(\phi')$ example of (\ref{eq:8}) obtains.
Put frivolously, their SAI equivalence shows that the two seemingly
far-reaching conclusions (\ref{eq:0a}) and (\ref{eq:0b}) are both in fact just
far-fetched.  

It seems that the authors of Ref.\ \cite{LSS} regard the SAI equivalence
of the two \emph{as-if} accounts as evidence in favor of the universality
they attribute to their ``classical random phase kicks'' picture. 
In fact, this wouldn't do justice to the enormous differences in the actual
physical situations. 
Despite the formal SAI equivalence of (\ref{eq:6})--(\ref{eq:8}) and 
(\ref{eq:9})--(\ref{eq:11}), the physical significance of the phases 
$\delta_\epsilon$ and $\delta_\epsilon+\pi$ in (\ref{eq:8}) is utterly
different from the one in~(\ref{eq:13}).
 
\subsection{Quantum erasure}\label{sec:QE}
If the transition (\ref{eq:5}) is actually the result of classical phase kicks
that are truly random, then the value of $\epsilon$ is all the experimenter
can ever know. In particular, he cannot determine the $\Omega$-weight of 
(\ref{eq:6}) and (\ref{eq:7}) because the random nature of the phase kicks 
precludes any further knowledge \cite{FN3}.

In marked contrast, much information is potentially available to the
experimenter if quantum entanglement causes the transition (\ref{eq:5}),
in particular the entanglement  with a well controlled which-way 
detection device. For instance, the experimenter could perform a judiciously 
chosen measurement on the final state of the device to learn about the way 
taken. More relevant for the present discussion is the possibility of quantum
erasure \cite{SD,ESWajp}: The experimenter sorts the interfering quantum
systems 
into subensembles in accordance with the outcome of a measurement that 
determines in which one of the $U_{\rm env}$ eigenstates the which-way 
detector is found. In the example of (\ref{eq:10}) and (\ref{eq:11}), the 
observable ${\ketbra{+}{+}-\ketbra{-}{-}}$ would be measured, for instance, 
and the outcomes $\pm1$ found with relative frequencies $(1\pm|\epsilon|)/2$.

The corresponding subensembles have statistical operators that project to
\begin{equation}\label{eq:15}
\ket{\psi^{(\pm)}}=\cos(\theta/2)\ket{\psi_1}
\pm\Exp{i(\phi+\delta_\epsilon)}\sin(\theta/2)\ket{\psi_2}\,,
\end{equation}
and each subensemble exhibits fringes with the original visibility 
$\visib_0=\sin\theta$.

If he'd wish to do so, the experimenter could then say that it is \emph{as if} 
the interaction had created two subensembles by the transitions
$\phi\to\phi+\delta_\epsilon$ and $\phi\to\phi+\delta_\epsilon+\pi$ and
allocated the fractions $(1\pm|\epsilon|)/2$ to them.
In this manner, the experimenter would actually determine $\Omega(\phi')$
of (\ref{eq:8}), and we emphasize that he has this possibility only if the
quantum-entanglement mechanism is actually at work. Quantum erasure is
impossible if the interfering system has suffered from truly random phase 
kicks.
Hence, the $\Omega$-weight has a phenomenological 
significance only if the ``quantum entanglement'' picture is actual, but not
when the ``classical random phase kicks'' are in action.

But, again there is an essential \emph{as-if} character to this ``interaction
creates subensembles'' notion. A measurement of another observable, such as
$\ketbra{0}{0}-\ketbra{1}{1}$, for instance, would result in a different pair
of subensembles with an equally justified \emph{as-if} interpretation. One of
the subensembles may yield fringes with a visibility $\visib_{\rm sub}$ that 
exceeds $\visib_0=\sin\theta$, even $\visib_{\rm sub}=1$ is achievable ---
an impossible feat if (\ref{eq:5}) resulted from ``classical random phase 
kicks.'' 

The quantum-erasure sorting into the subensembles of (\ref{eq:15}) labels
each 
of the interfering quantum systems by the phase ${+(\phi+\delta_{\epsilon})}$
or by the phase ${-(\phi+\delta_{\epsilon})}$, and as a consequence of the
quantum-mechanical indeterminacy, these labels are attached randomly. This
invites the question: `Aren't these labels just the classical random 
phases of Ref.\ \cite{LSS}?' Answer: No, they aren't, because the randomness
of the labeling process is of a purely \emph{quantum} nature and has
absolutely nothing to do with \emph{classical} phase noise. 

\section{Relevance for real-life experiments}
In closing, we'd like to remark on the relevance of these somewhat abstract
considerations for real-life laboratory experiments. In any interferometer
experiment, there are unavoidably parameters that cannot be fully controlled,
or maybe not at all. For example, the beam splitters and mirrors of a
Mach-Zehnder interferometer could be subject to erratic mechanical
vibrations,
and the resulting loss of fringe visibility would be correctly blamed on
``classical random phase kicks.'' Similarly, optical elements of modern atom 
interferometers \cite{Berman} employ laser beams, and their jitters give rise
to random phases. Deliberately introduced quantum entanglement comes in 
addition, so that a satisfactory description of an experiment must take both 
mechanisms for visibility loss into account.

The recent atom interferometer experiments by D\"urr, Nonn, and Rempe (DNR) 
\cite{DNR} illustrate these matters quite well. DNR had to be content with
a maximal fringe visibility of $\sim70\%$ although their interferometer is
symmetric [$\cos^2(\theta/2){\simeq\sin^2(\theta/2)}{\simeq\frac{1}{2}}$, 
${\sin\theta\lesssim1}$]. Clearly, lack of control over some parameters ---
the hallmark of \emph{random} noise --- is responsible for the ${\sim30\%}$
reduction. The which-way detection performed by DNR led to a complete
loss of fringe visibility --- undoubtedly the result of quantum 
\emph{entanglement}, as demonstrated by the resurrection of the fringes with
the aid of quantum erasure.


\begin{references}
\bibitem{FNx}
Incidentally we note that, for all values of the phase parameter $\varphi$,
the $I$ integration yields the projector to the corresponding London's phase
states 
[F. London, Z. Phys.\ {\bf 37}, 915 (1926); {\bf 40}, 193 (1927)].
\bibitem{FN1}
For simplicity we don't consider the additional possibility of randomizing
the angle parameter $\theta$ in (\ref{eq:1}).
\bibitem{FN2}
More general possibilities can be considered; for the sake of simplicity, 
we are content with (\ref{eq:12}).
\bibitem{FN3}
In Ref.\ \cite{LSS} the interference pattern (\ref{eq:4}) is regarded as
($\pi$ times) a probability distribution for the phase variable $\varphi$, and
then associated with a phase dispersion. 
This, however, has no bearing on the issue at hand, since
$p_\epsilon(\varphi)$ contains no information about  $\Omega(\phi')$ beyond
what is stated in Eqs.\ (\ref{eq:6}) and (\ref{eq:7}).
\end{references}
\end{document}